\documentclass[10pt]{iopart}
\usepackage{iopams}
\usepackage{bm}
\usepackage{amssymb}
\usepackage[numbers]{natbib}
\usepackage[a4wide]{}
\usepackage[latin1]{inputenc}
\usepackage[english]{babel}
\usepackage{graphicx}
\usepackage{textcomp}
\usepackage{subfigure}
\usepackage{lscape}
\usepackage{color}
\usepackage{verbatim} 
\usepackage{hyperref}   

  \expandafter\let\csname equation*\endcsname\relax
  \expandafter\let\csname endequation*\endcsname\relax
\usepackage{amsmath}

\def\ba{{\bm a}}
\def\bb{{\bm b}}

\def\bN{{\bm N}}

\def\bx{{\bm x}}

\def\bR{{\bm R}}

\def\rab{\mbox{ $R_{AB}$}}

\def\nn{\bN^{(1)}.\bN^{(2)}}

\def\ba{{\bm a}}
\def\bb{{\bm b}}

\def\bx{{\bm x}}

\def\bN{{\bm N}}
\def\bR{{\bm R}}

\def\rab{R_{AB}}

\def\lb{\label}
\def\be{\begin{equation}}
\def\ee{\end{equation}}
\def\bea{\begin{eqnarray}}
\def\eea{\end{eqnarray}}

\newcommand{\abs}[1]{\left | #1 \right |}

\def\lb{\label}

\def\be{\begin{equation}}
\def\ee{\end{equation}}
\def\bea{\begin{eqnarray}}
\def\eea{\end{eqnarray}}

\def\bx{{\bm x}}  
\def\ba{{\bm a}}  

\newcommand{\norm}[1]{|{#1}|}

\def\lb{\label}
\def\nab{N_{AB}}

\def\brpa{\bm{R_{PA}}}
\def\brpb{\bm{R_{PB}}}
\def\rpa{R_{PA}}
\def\rpb{R_{PB}}
\def\bbetap{\bm{\beta_P}}

\def\bnab{\bm N_{AB}}

\def\bg{\bm g_P}

\def\bR{\bm R}

\def\nn{\nonumber}

\def\be{\begin{equation}}
\def\ee{\end{equation}}
\def\bea{\begin{eqnarray}}
\def\eea{\end{eqnarray}}

\def\vp{\bm{v}_{P}}
\def\app{\bm{a}_{P}}

\def\mirdr{{\cal O}(\Delta_{x_P})}
\def\myrdr{{\cal O}(R_P \, \Delta_{x_P})}
\def\bR{\bm{R}}
\def\bN{\bm{N}}

\newcommand\mordre[1] {{\cal O}(c^{- {#1}})}

\begin{document}

\title[Time transfer functions to validate light propagation solutions for astrometry]{Time Transfer functions as a way to validate light propagation solutions for space astrometry}

\author{Stefano~Bertone$^{1,2}$, Olivier~Minazzoli$^{3}$, Mariateresa~Crosta$^2$, Christophe~Le~Poncin-Lafitte$^1$, Alberto~Vecchiato$^2$ and Marie-Christine~Angonin$^1$}
\address{$^1$ Observatoire de Paris, SYRTE,
CNRS/UMR 8630, UPMC \\
61 avenue de l'Observatoire, F-75014 Paris, France}
\address{$^2$ INAF, Astrophysical Observatory of Torino, University of Torino \\ Via Osservatorio 20, 10025 Pino Torinese (Torino), Italy }
\address{$^3$ UMR ARTEMIS, CNRS, University of Nice Sophia-Antipolis, Observatoire de la C\^ote d'Azur, BP4229, 06304, Nice Cedex 4, France}

\date{\today}

\begin{abstract} 
Given the extreme accuracy of modern space astrometry, a precise relativistic modeling of observations is required. Concerning light propagation, the standard procedure is the solution of the null-geodesic equations. However, another approach based on the Time Transfer Functions (TTF) has demonstrated its capability to give access to key quantities such as the time of flight of a light signal between two point-events and the tangent vector to its null-geodesic in a weak gravitational field using an integral-based method. 
The availability of several models, formulated in different and independent ways, must not be considered like an oversized relativistic toolbox. Quite the contrary, they are needed as validation to put future experimental results on solid ground.
The objective of this work is then twofold. First, we build the time of flight and tangent vectors in a closed form within the TTF formalism giving the case of a time dependent metric. Second, we show how to use this new approach to obtain a comparison of the TTF with two existing modelings, namely GREM and RAMOD. In this way, we evidentiate the mutual consistency of the three models, opening the basis for further links between all the approaches, which is mandatory for the interpretation of future space missions data. This will be illustrated through two recognized cases: a static gravitational field and a system of monopoles in uniform motion.
\end{abstract}


\maketitle

%%%%%%%%%%%%%%%%%%%%%%%%%%%%%%%%%%%%%%%%%%%%%%%%%%%%%%%%%%%%%%%%%%%%%%%%%%
\section{Introduction}
%%%%%%%%%%%%%%%%%%%%%%%%%%%%%%%%%%%%%%%%%%%%%%%%%%%%%%%%%%%%%%%%%%%%%%%%%%

Modern astrometry relies on high precision observations whose data need to be reduced and interpreted in the framework of General Relativity (GR)~\cite{1991AJ....101.2306S,Moyer:2000,1992AJ....104..897K,2003AJ....126.2687S,2003AJ....125.1580K,2004ApJ...607..580D}. To reach the demanded precision, several key points need to be considered: the definition of the observation in a proper reference frame, global reference systems allowing the comparison of observations made in each proper reference frame and a precise modeling for the propagation of the observed signal. Each of these issues has been deeply studied in the literature: the definition of global reference systems has been given by the IAU 2000 Resolution B1.3 in the post-Newtonian approximation of GR~\cite{2003AJ....126.2687S} while several relativistic definitions of physically adequate local reference frames of a test observer have been proposed in~\cite{2003CQGra..20.4695B,2004PhRvD..69l4001K}. As mentioned above, a precise modeling for the relativistic propagation of Electromagnetic Waves (EW) is also required. In fact, the behaviour of the EW in the Solar System is intrinsically related to space-time's curvature and therefore one has to take it into account for modern astrometry. For instance, the astrometric mission Gaia~\cite{2002EAS.....2.....B} is expected to reach an accuracy of several microarcseconds ($\mu as$) for the positions, parallaxes and proper motion of remote celestial sources while post-Newtonian corrections to light direction due to the gravitational field of Solar System's bodies can reach $16$~milliarcseconds~($mas$) for a light ray grazing Jupiter~\cite{2003AJ....125.1580K}. 

In this paper, we will focus on modelling the propagation of EW. In the paradigm of Maxwell electromagnetism minimally coupled to gravitation through the space-time metric, EW in their geometric optics limit are known to follow null geodesics~\cite{gravitationBook}. Assuming that the metric is known, solving the null geodesic equations is the standard method allowing to get all the information about light propagation between two point-events. Many solutions have been proposed in the post-Newtonian~(PN) and the post-Minkowskian~(PM) approximations when dealing with a metric tensor taking into account the dynamical behaviour of the Solar System~\cite{1992AJ....104..897K, 2007PhRvD..75f2002K,2011CQGra..28h5010M,2012PhRvD..86d4007D, 2006ApJ...653.1552D}. However, it has been demonstrated that solving the null geodesic equations is not mandatory and can be replaced by another approach based on the Time Transfer Functions (TTF)~\cite{2004CQGra..21.4463L,2008CQGra..25n5020T}. If the TTF approach does not provide the full trajectory of light, it gives a formulation in closed form of what is needed for space astrometry, namely the time of flight of an EW between two point-events and the tangent vector to the null-geodesic at the observation event. TTF have been formulated as a general post-Minkowskian series of ascending powers of the Newtonian gravitational constant $G$~\cite{2008CQGra..25n5020T}, which has not yet been done using null-geodesic approaches~; explicit solutions have been obtained and tested in two cases: the PN stationary axisymmetric gravitating body~\cite{2008PhRvD..77d4029L} and a static monopole body at second and third PM order of approximation~\cite{2002PhRvD..66b4045L,2010CQGra..27g5015K,2010CQGra..27n5013A, 2012CQGra..29x5010T,2012MmSAI..83.1024T}.  
It has been recently shown~\cite{2012MmSAI..83.1020B} that the TTF can also be applied to astrometry. 
However, a deeper study would be needed to apply this formalism to observations from within a realistic description of the Solar System.

At present time, two robust modelings have been developed for Gaia~: GREM~\cite{2003AJ....125.1580K} and RAMOD~\cite{2011CQGra..28w5013C}. Both are based on the solution of the null-geodesic equations even if starting from a different definition of the involved quantities. Briefly, GREM is formulated using a coordinate approach and the IAU reference systems, while RAMOD bases the ray-tracing problem on a measurement protocol~\cite{deFelice.book} to maintain the general relativistic conception of the involved unknowns.
Since they will operate on the same set of real data, it is fundamental to be able to compare them. 
From the experimental point of view, in fact, modern space astrometry is going to bring our knowledge into a widely unknown territory. Such a huge push-forward will not only come from high-precision measurements, which call for a suitable relativistic modeling, but also in form of absolute results which can hardly be validated by independent, ground-based observations. In this sense, it is of capital importance to have different, and cross-checked models to interpret these experimental data.

The goal of this paper is then twofold.
In its first part we will present a study of the TTF formalism in a dynamical case that is well suited to describe the Solar System at the needed accuracy~\cite{2012PhRvD..86d4007D}.
Second, we show how to use this new approach to obtain a consistency check with GREM and RAMOD on two well-known quantities, namely the time of flight and the tangent vectors to the null geodesic. To illustrate this, we consider both a static gravitational field and the case of monopoles in uniform motion. 

The paper is organized as follows. Section~\ref{sect:2} gives the notations used in this article. In section~\ref{sect:3} we give a short review of the TTF formalism in the post-Newtonian approximation while in section~\ref{sect:3a} we present a new method to obtain the tangent vectors in a closed form within the TTF formalism. The equations describing EW propagation in a dynamical system are then explicitly given in section~\ref{sect:4}.
Section~\ref{sect:5} shows the procedure to interface the geodesic approaches to the TTF and finally, in section~\ref{sect:concl} we give our concluding remarks.

%%%%%%%%%%%%%%%%%%%%%%%%%%%%%%%%%%
\section{Notation and conventions}  \label{sect:2}
%%%%%%%%%%%%%%%%%%%%%%%%%%%%%%%%%%
In this paper $c$ is the speed of light in a vacuum and $G$ is the Newtonian gravitational constant. The Lorentzian metric of space-time $V_4$ is denoted by $g$. The signature adopted for $g$ is $(-+++)$. We suppose that space-time is covered by a global quasi-Galilean coordinate system~\cite{lrr-2006-3} $(x^\mu)=(x^0,\bx )$, where $x^0=ct$, $t$ being a time coordinate, and $\bx=(x^i)$. We assume that $g_{00}<0$ anywhere. We employ the vector notation $\ba$ in order to denote $(a^1,a^2,a^3)=(a^i)$. 
Considering two such quantities $\ba$ and $\bb$ we use $\ba \cdot \bb$ to denote $a^i b^i$ (Einstein convention on repeated indices is used). The quantity $a=\vert \ba \vert$ stands for the ordinary Euclidean norm of $\ba$. 
For any quantity $f(x^{\lambda})$, $f_{, \alpha}$ and $\partial_{\alpha} f$ denote the partial derivative of $f$ with respect to $x^{\alpha}$. The indices in parentheses characterize the order of perturbation. They are set up or down, depending on the convenience. 

%%%%%%%%%%%%%%%%%%%%%%%%%%%%%%%%%%
\section{Time Transfer Functions formalism} \label{sect:3}
%%%%%%%%%%%%%%%%%%%%%%%%%%%%%%%%%%

In this section, we recall the basics and the properties of the~TTF formalism. This method stands as a development of Synge World Function~\cite{SyngeBookGR}, an integral approach based on the principle of minimal action (see~\cite{2008CQGra..25n5020T} and references herein) and containing all the informations about an EW. 
While the World Function is an implicit equation of the photon trajectory nearly impossible to solve, the TTF formalism gives up some generality to provide important information about the propagation of an EW between two points at finite distance: the time of flight $\mathcal{T}_{e/r}$ which is important in various fields of astronomy and space science, such as the positioning of space probes or the lunar laser ranging; the knowledge of the tangent vectors to the light ray, obviously required for astrometry; the frequency shift of a signal between two points, which has applications in many fields of space science.
The reader can refer to~\cite{2012MmSAI..83.1020B,2012CQGra..29w5027H,2009A&A...499..331L} and references herein for more details. 

Let us define $x_A = (ct_A, \bx_A)$ the event of emission ${\cal A}$ and $x_B = (ct_B, \bx_B)$ the event of reception ${\cal B}$ of a light signal. We denote ${\cal T}_{e}$ and ${\cal T}_{r}$ as two distinct (coordinate) time transfer functions defined as  
\begin{equation}  \label{ttf}
t_B - t_A = {\mathcal T}_e(t_A, \bx_A, \bx_B) = {\mathcal T}_r(t_B, \bx_A, \bx_B) \, ,
\end{equation}
where $\mathcal T_e$ and $\mathcal T_r$ are evaluated at the event of emission ${\cal A}$ and at  the event of reception ${\cal B}$, respectively. 

We shall consider a weak gravitational field so that we can write
\begin{equation}
g_{\mu\nu}=\eta_{\mu\nu}+h_{\mu\nu}\, ,
\end{equation}
with $\eta_{\mu \nu}=diag(-1,+1,+1,+1)$ the Minkowskian background and $h_{\mu \nu}$ a small perturbation. The general PM expansion of this formalism has been given in~\cite{2008CQGra..25n5020T} but in this work we shall consider only the slow-motion, post-Newtonian approximation~\cite{1993tegp.book.....W} - a case well adapted to our Solar System in which the two approximations coincide. So, we assume that the potentials $h_{\mu \nu}$ may be expanded as~\cite{2011CQGra..28h5010M}
\bea \lb{1n}
& & h_{00}=\frac{1}{c^2}h_{00}^{(2)}+
{\cal O}\left(\frac{1}{c^4}\right) \nonumber \, , \\
& & h_{0i}=\frac{1}{c^3}h_{0i}^{(3)}+{\cal O}\left(\frac{1}{c^4}\right) \, , \\ 
& & h_{ij}=\frac{1}{c^2}h_{ij}^{(2)}+{\cal O}\left(\frac{1}{c^4}\right) \, . \nonumber
\eea

Under these hypothesis, the time of flight $\mathcal{T}_{e/r}$ of a photon between $x_A$ and $x_B$ is given by the expressions~\cite{2002PhRvD..66b4045L,2008ASSL..349..153T}
\begin{subequations}  \lb{ttf_order}
\bea
	{\cal T}_{	r}(\bx_A, t_B, \bx_B)&=&\frac{R_{AB}}{c}+ \frac{1}{c} \Delta_{r}(\bx_A, t_B,\bx_B) + {\cal O}(c^{-5})  \, , \\ 
	{\cal T}_{	e}( t_A, \bx_A, \bx_B)&=&\frac{R_{AB}}{c}+ \frac{1}{c} \Delta_{e}(t_A, \bx_A, \bx_B) + {\cal O}(c^{-5})  \, ,
\eea
\end{subequations}
where $R_{AB} \equiv \vert {\bR_{AB}} \vert$ with $\bR_{AB} \equiv \bx_B-\bx_A$; $\Delta_{e/r}$ are the so called "delay functions"~\cite{2008CQGra..25n5020T} and represent the gravitational delay in the time of flight of the photon with respect to the Newtonian time of flight, defined as~\footnote{with the signature recommended by the IAU \citep{2003AJ....126.2687S} $(- + + +)$ and whose contravariant form $h^{\mu\nu}$ with signature $(+ - - -)$ is given in~\citep{2004CQGra..21.4463L} as a post-Minkowskian expansion.}
\begin{subequations}  \lb{Tr1PN}
\bea
\!\!\!\!\!\!\!\!\!\!\!\! &&\Delta_{r} = \frac{R_{AB}}{2 c^2} \int_{0}^{1}\left[h^{(2)}_{00} + \frac{2}{c} N_{AB}^{i} h^{(3)}_{0i} + N_{AB}^{i} N_{AB}^{j} h^{(2)}_{ij}\right]_{z^\alpha_{-}(\lambda)} d\lambda \, ,\lb{Tr1wIAUs}  \\
\!\!\!\!\!\!\!\!\!\!\!\! &&\Delta_{e} = \frac{R_{AB}}{2 c^2} \int_{0}^{1}\left[h^{(2)}_{00} + \frac{2}{c} N_{AB}^{i} h^{(3)}_{0i} + N_{AB}^{i} N_{AB}^{j} h^{(2)}_{ij}\right]_{z^\alpha_{+}(\mu)} d\mu \, ,\lb{Te1wIAUs}  
\eea
\end{subequations}
with $\bN_{AB} \equiv \dfrac{\bR_{AB}}{R_{AB}}$.
The two integrals are taken along the Minkowskian paths $z^\alpha_{-}(\lambda) = (x^0_B - \lambda R_{AB}, x^i_B - \lambda R^i_{AB})$ and $z^\alpha_{+}(\mu) = (x^0_A + \mu R_{AB}, x^i_A + \mu R^i_{AB})$, which represent the unperturbed "straight lines" between $x_A$ and $x_B$, respectively.
\noindent   
%%%%%%%%%%%%%%%%%%%%%%%%%%%%%%%%%%
\section{Tangent vectors in closed form} \label{sect:3a}
%%%%%%%%%%%%%%%%%%%%%%%%%%%%%%%%%%
The TTF formalism also provides a direct way of defining the tangent vectors of a photon $k^{\mu} \equiv dx^{\mu}/d\lambda$ at its reception event, as follows 
\begin{eqnarray}  \lb{eq:k}
\!\!\!\!\!\!\!\!& &\left(\widehat{k}_i\right)_B \equiv \left(\frac{k_i}{k_0}\right)_B =
-c \, \frac{\partial {\cal T}_{e}}{\partial x^{i}_{B}} \, = \,
- c \, \frac{\partial  {\cal T}_{r}}{\partial x^{i}_{B}}
\left[1 - \frac{\partial  {\cal T}_{r}} {\partial t_B}\right]^{-1} \; .  \label{2d1}
\end{eqnarray} 
Similarly, one can define the tangent vector $\left(\widehat{k}_i\right)_A$ at emission event as shown in~\cite{2004CQGra..21.4463L}. 

We can outline the following procedure. Let us expand Eq.\eqref{eq:k} as function of the gradient of the delay functions $\Delta_{e}$ and $\Delta_{r}$ using Eq.~\eqref{ttf_order}
\bea \lb{eq:kCF}
\fl	\left(\widehat{k}_i\right)_B &=& N^i_{AB} + \frac{\partial \Delta_{e}}{\partial x^i_{B}} + {\cal O}(c^{-5}) = N^i_{AB} + \frac{\partial \Delta_{r}}{\partial x^i_{B}} + N^i_{AB} \frac{\partial \Delta_{r}}{\partial x^0_{B}} + {\cal O}(c^{-5}) \; .
\eea
Then, Eq.~\eqref{Tr1PN} allows us to express the gradients of $\Delta_{e}$ and $\Delta_{r}$ as integral functions of the metric and its derivatives in order to build up the definition of the tangent vectors in a closed form.
By defining $(\nab)^i = \delta^{ij} (\nab)_j$ and
\begin{subequations}
\bea
	m_{,\alpha} &\equiv& h_{00,\alpha} + 2 N_{AB}^k h_{0k,\alpha} + N_{AB}^j N_{AB}^k h_{jk,\alpha} \; , \\
	\tilde h_i &\equiv& N_{AB}^i h_{00} - N_{AB}^i N_{AB}^j N_{AB}^k h_{jk} + 2 h_{0i} + 2 N_{AB}^j h_{ij} \quad , 
\eea
\end{subequations}
the gradients appearing in Eq.\eqref{eq:kCF} can be computed as
\begin{subequations}  \lb{eq:dDr1PM}
\bea
\fl	\frac{\partial \Delta_r }{\partial x^i_B} &=& -\frac{1}{2} \int_0^1 \left[  R^i_{AB} \lambda m_{,0} - R_{AB} (1-\lambda) m_{,i} - \tilde h_i  \right]_{z_- (\lambda)} d\lambda \; , \\
\fl	\frac{\partial \Delta_r }{\partial x^0_B} &=& \frac{R_{AB}}{2} \int_0^1 \left[ m_{,0}  \right]_{z_-(\lambda)} d\lambda \; , \\
\fl	\frac{\partial \Delta_e }{\partial x^i_B} &=&  -\frac{1}{2} \int_0^1 \left[  - R^i_{AB} \mu m_{,0} - R_{AB} \mu m_{,i} + \tilde h_i \right]_{z_+ (\mu)} d\mu  \; .
\eea
\end{subequations}

%%%%%%%%%%%%%%%%%%%%%%%%%%%%%%%%%%%%%%%%%%%%%%%%%%%%%%%%%%%%
\section{Time transfer and light propagation in the Solar System} \lb{sect:4}
%%%%%%%%%%%%%%%%%%%%%%%%%%%%%%%%%%%%%%%%%%%%%%%%%%%%%%%%%%%%

We provide now explicit equations for the TTF formalism presented in sections~\ref{sect:3}~and~\ref{sect:3a} for the case of point-like, slowly moving and non-rotating bodies~\footnote{we neglect here the effects due to the asphericity of the field sources since its contribution is already given in~\cite{2008PhRvD..77d4029L} for the stationary case and its extension to moving sources is irrelevant at the level of precision of most experiments at present time.}.
This system can be represented by a PPN metric tensor admitting the following perturbation $h_{\mu \nu}$
\be \lb{IAUmetric}
\fl	\qquad \qquad h_{00}= \frac{2G}{c^2} \sum_P \frac{ {\cal M}_P }{R_P(t,\bx)} \;, \qquad	h_{0i}= - (1+\gamma) h_{00} \beta^i_P (t) \;, \qquad	h_{ij}= \delta_{ij} \gamma h_{00} \;, 
\ee
with ${\cal M}_P$ the mass of the perturbing body $P$ and $R_P$ its distance to the photon trajectory $\bm{x}(t)$ at coordinate time $t$; $\beta^i_P (t) = \dfrac{v^i_P (t)}{c}$ is the ratio of the perturbing body barycentric velocity at coordinate time $t$ to the speed of light $c$ and $\gamma$ is a PPN parameter~\cite{1993tegp.book.....W}. 

Following the usual assumption~\cite{1992AJ....104..897K} regarding the trajectory of the perturbing bodies $\bx_P$, we consider that they are rectilinear and uniform so that
\be
\bx_P(t)=\bx_P(t_{C})+c(t-t_{C}) \bm{\beta}_P (t_C) +\mirdr\; ,  \lb{eq:dev_x}
\ee
where $\Delta_{x_P}$ is some typical error made on the position of the perturbing body due to the linear approximation chosen for its trajectory and below the desired accuracy of our model (see~\ref{sec:approxtrajP} for more details) and $t_{C}$ is some fixed moment of time detailed in~\ref{sec:tDprob}. This choice will also be useful in section~\ref{sect:5} when comparing our results to other astrometric modelings. 

%%%%%%%%%%%%%%%%%%%%%%%%%%%%%%%%%%%%%%%%%%%%%%%%%%%%%%%%
\subsection{Time transfer functions in the case of moving monopoles} \lb{TTFmovmon}
%%%%%%%%%%%%%%%%%%%%%%%%%%%%%%%%%%%%%%%%%%%%%%%%%%%%%%%%
Taking into account Eq.~\eqref{IAUmetric}, Eq.~\eqref{Tr1PN} writes at first order
\begin{subequations}  \lb{eq:TTFdelay}
\bea
\fl \Delta_{r}^{(1)}(\bx_A, t_B, \bx_B) &=& (\gamma+1) R_{AB}  \frac{G}{c^2} \sum_P {\cal M}_P g_P^2 \int_{0}^{1} \left[\frac{1}{R_P(t,\bx)} \right]_{z^\alpha_-(\lambda)} d\lambda,   \\
\fl \Delta_{e}^{(1)}(t_A, \bx_A, \bx_B) &=& (\gamma+1) R_{AB} \frac{G}{c^2} \sum_P {\cal M}_P g_P^2 \int_{0}^{1} \left[\frac{1}{R_P(t,\bx)} \right]_{z^\alpha_+(\mu)} d\mu    ,
\eea
\end{subequations}
where we defined $g^i_P \equiv N^i_{AB} - \beta^i_P (t_C)$. Remembering the definition 
\be \lb{eq:RPbase}
R^2_P(t)=\left[ \bm{x}(t)-\bm{x}_P(t) \right]^2 \; ,
\ee
where $\bm{x}(t)$ stands for the coordinate position along the integration path, we can expand it using Eq.~\eqref{eq:dev_x}
\bea \lb{eq:RPcompl}
\fl R^2_P(t,\bx)|_{z_-^\alpha(\lambda)}&=&\big[ \bm{x}(t)-\bm{x}_P(t)  \big]^2|_{z_-^\alpha(\lambda)}  \nn \\
\fl &=&  \left[\bm{x}_P(t_{C})+ c~(t-t_{C})|_{z_-^0(\lambda)} \bm{\beta}_P (t_C) - \bm{x}_B + \lambda \bm{R}_{AB}  \right]^2 +\myrdr\nonumber \\
\fl &=& \Big[\brpb - \lambda \rab \bg \Big]^2 +\myrdr  
\eea
or
\bea
\fl R^2_P(t,\bx)|_{z_+^\alpha(\lambda)}&=&\big[ \bm{x}(t)-\bm{x}_P(t)  \big]^2|_{z_+^\alpha(\mu)}   \nn \\
\fl &=&  \left[ \bm{x}_P(t_{C})+ c~(t-t_{C})|_{z_+^0(\mu)} \bm{\beta}_P (t_C) - \bm{x}_A - \mu \bm{R}_{AB}  \right]^2 +\myrdr \nonumber \\
\fl &=& \Big[\brpa + \mu \rab \bg \Big]^2 +\myrdr, 
\eea
where for practical reasons we set the notation
\begin{subequations} \lb{eq:RPdefs}
\bea 
&&\bm{R}_{PX}=\bx_X - \bx_P(t_{C}) + c(t_X - t_C) \bm{\beta}_P (t_C) ,  \\
&&  R_{PX}=|\bm{R}_{PX}|, \qquad \bm{N}_{PX} = \frac{\bm{R}_{PX}}{R_{PX}} \; ,   \\
&&\bm{R}_{AB}=\bx_B-\bx_A,~~R_{AB}=|\bm{R}_{AB}|,~~\bm{N}_{AB} = \frac{\bm{R}_{AB}}{R_{AB}} \; . 
\eea
\end{subequations}
Noting the boundary conditions 
\begin{subequations} \lb{eq:rel_R_P}
\bea 
	\bm R_P (0) &=& \bm R_{PB}  \; ,\\
	\bm R_P (1) &=& \bm R_{PB} - R_{AB} \bm g_P \equiv \bm R_{PA} \; ,\\
	\bm R_{PB} - \bm R_{PA} &=&  \bm g_P R_{AB}  
\eea
\end{subequations}
and substituting for $R_P$ from Eqs.~\eqref{eq:RPcompl}-\eqref{eq:RPdefs} into Eq.\eqref{eq:TTFdelay}, after some algebra one gets the reception and emission delay functions $\Delta^{(1)}_r(\bx_A, t_B, \bx_B)$ and $\Delta^{(1)}_e (t_A, \bx_A, \bx_B)$ as functions of the reception/emission coordinates
\bea \lb{ttf_ordre1}
\fl\Delta_{r}^{(1)} &=&  (\gamma+1) \frac{G}{c^2} \sum_P {\cal M}_P \left( \bg \cdot \bnab \right) \ln \left[ \frac{ g_P \rpa -  \brpa \cdot \bg }{ g_P \rpb - \brpb \cdot \bg} \right]   \\
\fl&=& (\gamma+1) \frac{G}{c^2} \sum_P {\cal M}_P \Big[ 1 - \bbetap (t_C) \cdot \bnab \Big] \nonumber\\
\fl&&\qquad\qquad\;\times\ln \left[ \frac{\rpa - \brpa \cdot \bnab - \bbetap (t_C) \cdot ( \brpa -  \bnab \rpa)}{\rpb - \brpb \cdot \bnab - \bbetap (t_C) \cdot ( \brpb - \bnab \rpb)} \right] \nn \\
\fl &=& (\gamma+1) \frac{G}{c^2} \sum_P {\cal M}_P \Bigg\lbrace \ln \left( \frac{\rpa - \brpa \cdot \bnab}{\rpb - \brpb \cdot \bnab} \right) \nonumber\\
\fl&& \quad\quad\quad\qquad+ \bbetap (t_C) \cdot \left[ \bnab \ln \left( \frac{\rpa - \brpa \cdot \bnab}{\rpb - \brpb \cdot \bnab} \right) \right. + \nn \\ 
\fl	&&	\quad\quad\quad\qquad\qquad\qquad\quad\left. \frac{\brpb -  \bnab \rpb}{\rpb-\brpb \cdot \bnab} - \frac{\brpa -  \bnab \rpa}{\rpa-\brpa \cdot \bnab} \right]\Bigg\rbrace \nn 
\eea
and
\bea \lb{ttf_ordre1_e}
\fl  \Delta_{e}^{(1)} &=& (\gamma+1) \frac{G}{c^2} \sum_P {\cal M}_P \ln \left[ \frac{ g_P \rpb + \brpb \cdot \bg}{ g_P \rpa +  \brpa \cdot \bg} \right]\nonumber\\
\fl			&=& (\gamma+1) \frac{G}{c^2} \sum_P {\cal M}_P \Big[ 1 - \bbetap (t_C) \cdot \bnab \Big] \\
\fl &&\qquad\qquad\;\times\ln \left[ \frac{\rpb + \brpb \cdot \bnab - \bbetap (t_C) \cdot ( \brpb +  \bnab \rpb)}{\rpa + \brpa \cdot \bnab - \bbetap (t_C) \cdot ( \brpa +  \bnab \rpa)} \right]  \nn \\
\fl			&=& (\gamma+1) \frac{G}{c^2} \sum_P {\cal M}_P \Bigg\lbrace \ln \left( \frac{\rpb + \brpb \cdot \bnab}{\rpa + \brpa \cdot \bnab} \right) \nonumber\\
\fl&& \quad\quad\quad\qquad+ \bbetap (t_C) \cdot \left[ \bnab \ln \left( \frac{\rpb + \brpb \cdot \bnab}{\rpa + \brpa \cdot \bnab} \right) \right. + \nn \\ 
\fl&& \quad\quad\quad\qquad\qquad\qquad\quad\left. \frac{\brpa +  \bnab \rpa}{\rpa+\brpa \cdot \bnab} - \frac{\brpb +  \bnab \rpb}{\rpb+\brpb \cdot \bnab} \right]\Bigg\rbrace. \nn
\eea
By setting $\bbetap = 0$ and $\bg = \bnab$ in Eq.~\eqref{ttf_ordre1}, we retrieve the static case given in~\cite{2002PhRvD..66b4045L}. Moreover, in~\ref{appendix1} we demonstrate the equality $\Delta_{r}^{(1)} = \Delta_{e}^{(1)}$ as required by Eq.~\eqref{ttf}.
Finally, we also applied Eq.~\eqref{ttf_ordre1} to the simple configuration of a signal propagating from the outer Solar System to the Earth and grazing Jupiter. Our evaluation of the gravito-electric field caused by the orbital motion of Jupiter on the time of flight of the photon is of the order of $10~ps$, in accordance to previous results~\cite{2002PhRvD..66b4045L}.

%%%%%%%%%%%%%%%%%%%%%%%%%%%%%%%%%%%%%%%%%%%%%
\subsection{Tangent vectors in the case of moving monopoles} 
%%%%%%%%%%%%%%%%%%%%%%%%%%%%%%%%%%%%%%%%%%%%%

We provide here the steps to compute the tangent vector at reception event $\left(\widehat{k}_i\right)_B(\bx_A, t_B, \bx_B, \bx_P, \bm \beta_P, \gamma)$. From Eq.~\eqref{eq:kCF}, Eq.~\eqref{eq:dDr1PM}, with the metric~\eqref{IAUmetric} and the notations~\eqref{eq:RPbase}-\eqref{eq:rel_R_P}, we first need to compute the partial derivatives of~$h_{00} (x,t)$ as follows
\be \lb{eq:hoopartder}
\fl \qquad \qquad	h_{00,i} = - \frac{2 G}{c^2} \sum_P {\cal M}_P \frac{R_P^i}{R^3_P} \; , \qquad	h_{00,0} = \frac{2 G}{c^2} \sum_P {\cal M}_P \frac{\bm R_P \bm  \cdot \bbetap (t_C) }{R^3_P} \; .
\ee
Using now the results of sections~\ref{sect:3a}, \ref{TTFmovmon} and Eq.~\eqref{eq:hoopartder}, it yields the integral equation for the tangent vector
\bea \lb{eq:kintform}
\fl  && \left(\widehat{k}_i\right)_B = - \nab^i + (\gamma+1) \frac{G}{c^2} \sum_P {\cal M}_P \int_0^1 \left\{ R_{AB} g_P^2 \left[ \Big(  R_{AB} \beta_P^i (t_C) - R^i_{AB} g_P \Big)\frac{\lambda - \lambda^2}{R_P^3(\lambda)}  \right. \right. \nn \\
\fl			&&\qquad\qquad \quad \quad  \left. \left.  + \Big( R^i_{PB} - \nab^i \bm R_{PB} \cdot \bm{\beta}_P (t_C) \Big) \frac{1-\lambda}{R^3_P(\lambda)}  \right] +  \frac{2 \beta_P^i (t_C) - \nab^i}{ R_P(\lambda)} \right\} d\lambda \; ,
\eea
where the terms $\bm \beta$ and $\bg$ describe the deflection due to the dynamics of the system.

The explicit computation of the integrals appearing in the right-hand side (r.h.s.) of Eq.~\eqref{eq:kintform} may be obtained by taking into account the boundary conditions set in Eq.~\eqref{eq:rel_R_P}. After some algebra, we get
\bea \lb{eq:ki_monopmvmt}
\fl	\left(\widehat{k}_i\right)_B &=& - \nab^i + (\gamma+1) \frac{G}{c^2} \sum_P \frac{{\cal M}_P}{R_{AB} R_{PB} \Big[R_{PB}^2 g_P^2 - (\bm{R}_{PB} \cdot \bm{g}_P)^2 \Big] } \\
\fl				&&  \nn \\
\fl				&& \times \Bigg\{  g_P \nab^i \Big[ \Big( \bm{R}_{PB} \cdot \bm{N}_{AB} \Big) \Big(R_{PB}^2 - R_{PA} R_{PB} - R_{AB} \bm{R}_{PB} \cdot \bm{\beta}_P (t_C) \Big) \nn \\
\fl 				&&  \qquad\qquad -  R_{PB}^2 R_{AB} g_P^2 \Big] \nn \\
\fl				&&  \nn \\
\fl				&& \left. + R^i_{PB} g_P^2 \Big[ R_{PB} R_{PA} - R_{PB}^2 + R_{AB} \bm{R}_{PB} \cdot \bm{g}_P \Big]  \right. \nn \\
\fl				&&  \nn \\
\fl				&&  +\beta^i_P (t_C) R_{PB} \Big[ (R_{PA} - R_{PB}) (\bm R_{PB} \cdot \bm N_{AB}) + R_{PB} R_{AB} \Big]  \Bigg\} \nn \\
\fl				&&  \nn \\
\fl			   	&& + (\gamma+1) \frac{G}{c^2} \sum_P {\cal M}_P \frac{\beta_P^i (t_C) - \nab^i \bm{\beta}_P (t_C) \cdot \bm{N}_{AB} }{ R_{AB} g_P} \ln \frac{g_P R_{PB} + \bm R_{PB} \cdot \bm g_P}{g_P R_{PA} + \bm R_{PA} \cdot \bm{g}_P} \nn \\
\fl				&& + {\cal O}(c^{-4}) \; . \nn 
\eea
We shall note that, from the point of view of the astrometric data analysis, the last equation is obtained as a function of all known quantities ($i.e.$ the coordinates of the observing satellite and the mass distribution in the Solar System) and of the astrometric unknown ($i.e.$ the source coordinates).
By setting $\bbetap = 0$ and $\bg = \bnab$, the perturbing bodies are fixed at their position at time $t_{C}$ and we easily retrieve the static case~\cite{2008CQGra..25n5020T}.
It is also interesting to evaluate the gravito-electric contribution to the direction of light using the definition given in~\cite{2012CQGra..29x5010T}
\be
	\Delta \chi \approx \norm{\bnab \times \hat k_B} \; ,
\ee
where the light ray is considered as coming from infinity.
The expression of $ \hat k_B $ is then deduced from Eq.~\eqref{eq:ki_monopmvmt} where $\bnab \equiv \bm N$ and $\brpa \approx - \bm \rab$ in this case. Introducing the impact parameter $b_P$ and the angle $\alpha$ between $\bR_{PB}$ and $\bN$, we get $b_P = R_{PB} \sin{\alpha}$ so that
\bea \lb{eq: delta_infty}
\fl	&& \Delta \chi = (\gamma+1) \frac{G}{c^2} \sum_P \frac{{\cal M}_P}{R^2_{PB} \Big[ g_P^2 - (\bm{N}_{PB} \cdot \bm{g}_P)^2 \Big] } \bigg\{  b_P g^2_P \Big[1 + \bN_{PB} \cdot \bm g_P \Big]   \nn \\
\fl				&&  \nn \\
\fl				&& \qquad\qquad\qquad\qquad + \abs{\bN \times \bbetap} R_{PB} \big( 1-\bN_{PB} \cdot \bN \big) \bigg\} +  {\cal O}(c^{-4},R^{-1}_{AB}) \; .
\eea 
The logarithmic term disappears in Eq.~\eqref{eq: delta_infty} and can thus be neglected for sources at quasi-infinity. Moreover, numerical estimates of Eq.~\eqref{eq: delta_infty} for various deflecting Solar System bodies are in agreement with~\cite{2003AJ....125.1580K}.

%%%%%%%%%%%%%%%%%%%%%%%%%%%%%%%%%%%%%%%
\section{Relativistic astrometric models at the cross-checking point}\lb{sect:5}%How to compare different models of relativistic light propagation}\lb{sect:5}
%%%%%%%%%%%%%%%%%%%%%%%%%%%%%%%%%%%%%%%
The astrometric core solution of the forthcoming Gaia mission~\cite{2002EAS.....2.....B} will be performed by the Astrometric Global Iterative Solution (AGIS) software~\cite{2012A&A...538A..78L}. At the same time, an independent verification unit for AGIS called Global Sphere Reconstruction (GSR)~\cite{2012SPIE.8451E..3CV} has been set within the Gaia Data Processing and Analysis Consortium (DPAC). 
Since both pipelines are intended to operate on the same real data, the comparison of their results is mandatory in order to validate the final astrometric catalog.
In order to keep the two software as separate as possible, two different relativistic modelings of light propagation have been implemented: AGIS relies on GREM~\cite{2003AJ....125.1580K},  while GSR implements RAMOD~\cite{2006ApJ...653.1552D}. Both are derived from the solution of the null-geodesic equations but they present substantial differences in their formulation, in particular concerning the relativistic description of the involved quantities~\cite{2010A&A...509A..37C}. 
%It is then important to have a full procedure of comparison between them.
%As a third independent approach to relativistic astrometry, the TTF formalism offers the opportunity for a deeper analysis and cross-checking of the results.
%We therefore present a procedure to compare each of these models with the TTF, and as a consequence with each other as well. 
%This offers the opportunity for a further analysis and cross-validation of the results obtained within these three frameworks.
%As far as we know, such a comparison with the TTF has been done only for specific static model cases up to the second Post-Minkowskian order~\cite{2008CQGra..25n5020T,2010IAUS..261..103T}. 

In the following, we shall provide an analytical comparison of the TTF with KK92~\cite{1992AJ....104..897K}, a seminal study setting the basis for GREM,  and with RAMOD.
This comparison validates the results obtained in the previous section of this paper providing at the same time a cross-checking of these three relativistic models.

%%%%%%%%%%%%%%%%%%%%%%%%%%%%%%%%%%%%%%
\subsection{KK92 modeling} \lb{sec:KK92}
%%%%%%%%%%%%%%%%%%%%%%%%%%%%%%%%%%%%%%
KK92 describes light propagation in a gravitational system close to the one described by the metric assumed in Eq.~\eqref{IAUmetric}. Considering only the terms relevant for our purpose and using our notation, the trajectory of the photon can be written as
\be \lb{phot_traj}
	x^i (t) =  x^i (t_B) + c \sigma^i (t-t_B) + \Delta x^i  (t, x^i, t_B, x^i_B) \; ,
\ee
where $(t_B,x^i (t_B))$ are the reception coordinates, $\bm \sigma$ is a normalized vector giving the unperturbed direction of light at past null infinity and the gravitational perturbation is given by
\bea \lb{eq:deltax}
\fl \Delta x^i &=& - \frac{2G}{c^{2}} \sum_P \mathcal{M_P} 
\left \{ g^i_P \ln \left[\frac{\bm{g}_P \cdot \bm R_P(t) + g_P R_P(t)}{\bm{g}_P \cdot \bm R_{PB} + g_P R_{PB} } \right] + \frac{(\bm \sigma \times \bm R_P(t) \times \bm g_P)^i}{ g_P R_P(t) - \bm{g}_P \cdot \bm R_P(t)} \right.  \nonumber \\
\fl&& \phantom{-\frac{2G}{c^{2}} \sum_P \mathcal{M_P} } \left.- \frac{(\bm \sigma \times \bm R_{PB} \times \bm g_P)^i}{ g_P R_{PB} - \bm{g}_P \cdot \bm R_{PB}} \right\} \; . \qquad
\eea

The TTF formalism being designed for light propagation between two points located at finite distance, one has first to set the boundary condition 
\be \lb{eq:cross_eq}
	\bx(\bx_B,\bm \sigma,\Delta t)=\bx_A \\
\ee
in Eq.~\eqref{phot_traj} to provide the ''crossing trajectory equation''  
\be \lb{eq:trajAB}
	x^i (t_A) =  x^i (t_B) - c \, \Delta t \, \sigma^i + \Delta x^i  (\Delta t, x^i_B, \sigma^i) \; ,
\ee
where $\Delta t \equiv t_B-t_A$ represents the lapse of coordinate time between the emission and reception of the signal.
In the following, Eqs~\eqref{phot_traj}-\eqref{eq:trajAB} will be used to find the equivalence of KK92 and TTF for the time of flight and the tangent vectors.

%%%%%%%%%%%%%%%%%%%%%%%%%%%%%%%%%%%%%%
\subsubsection{Time of flight.} \lb{sect:ftKK92}
Let us state the formal development  
\be \lb{eq:deltatPN}
	{\Delta t} = \sum_i {\Delta t}_{(i)} \; , 
\ee
where ${\Delta t}_{(n)}$ is of order $\mordre{n}$. Substituting for $\Delta t$ from Eq.~\eqref{eq:deltatPN} into Eq.~\eqref{eq:trajAB} and identifying terms of the same order, we find 
\begin{subequations} \lb{eq:Ti}
	\bea 
		{\Delta t}_{(1)} &=& \frac{\rab}{c} \lb{geod_inv_0}  \; , \\ 
		{\Delta t}_{(2)} &=& \bnab \cdot \bm{\Delta x} (\Delta t, x^i_B, \sigma^i) \lb{geod_inv_1}  \\
				&=&  - \frac{2G}{c^2} \sum_P {\cal M}_P \left( \bg \cdot \bnab \right) \ln \left[\frac{\bm{g}_P \cdot \bm R_{PA} + g_P  R_{PA}}{\bm{g}_P \cdot \bm R_{PB} + g_P R_{PB} } \right]  \; , \nn
	\eea
\end{subequations}
where we used the property $\bm \sigma \cdot \bm \sigma = 1$ and noted that $\bnab \cdot (\bnab \times \bm R_X \times \bm g_P)=0$.
Using Eq.~\eqref{eq:equivAppC} shows that Eq.~\eqref{eq:Ti} is strictly equivalent to Eq.~\eqref{ttf_order} when the gravitational delay is given by Eq.~\eqref{ttf_ordre1} with $\gamma=1$. 

%%%%%%%%%%%%%%%%%%%%%%%%%%%%%%%%%%%%%
\subsubsection{Tangent vectors.}
The relation between the tangent vectors $k^\mu = \dfrac{dx^\mu}{d \lambda}$ and the photon velocity $\dot x^i$ used in KK92 is obtained by $\dfrac{\dot x^i}{c} = \dfrac{dx^i/d \lambda}{dx^0/d \lambda} = \dfrac{k^i}{k^0}$. It follows that
\bea \lb{eq:reldotxhatk}
	\hat k_i &=& \frac{k_i}{k_0} = \frac{g_{ij}k^j + g_{0i}k^0} {g_{00}k^0 + g_{0i}k^i} = \big( g_{0i} + g_{ij} \hat k^j \big) \big( g_{00} + g_{0i} \hat k^i \big)^{-1} \nn \\
		&=& - \frac{\dot x^i}{c} - 2 h_{00} \sigma^i - (\delta_{ij} + \sigma^i \sigma^j) h_{0j} + {\cal O} (c^{-4}) \; . \qquad
\eea

The computation of $\dfrac{\dot x^i}{c}$ is obtained by deriving the photon trajectory in Eq.~\eqref{phot_traj} with respect to coordinate time. Its application at $(t_B,\bx_B)$ gives 
\be \lb{eq:dotx}
	\frac{\dot x^i_B}{c} =  \sigma^i + \frac{\Delta \dot x^i_B}{c}  \; ,
\ee
where $\Delta \dot x^i $ represents the gravitational perturbation to the photon direction
\bea\lb{eq:deltaxi}
	\frac{\Delta \dot x^i_B}{c} &=& - \frac{2G}{c^{2}} \sum_P \mathcal{M}_P \frac{g_P}{\rpb}
\left \{ g_P^i + \frac{(\bnab \times \brpa \times \bg)^i}{g_P \rpb - \bg \cdot \brpb}  \right\} \; .  \qquad
\eea
Let us state the formal development  
\be \lb{eq:deltasigmaPN}
	\bm \sigma~=~\sum_i \bm \sigma_{(i)} \; , 
\ee
where $\bm \sigma_{(i)}$ is of order $\mordre{n}$. Substituting for $\bm \sigma$ from Eq.~\eqref{eq:deltasigmaPN} into Eq.~\eqref{eq:trajAB} and identifying all terms of the same order, we find
\begin{subequations} \lb{eq:sigmai}
	\bea 
		\sigma^i_{(1)} &=& \frac{x^i_B - x^i_A}{\rab} = N^i_{AB}  \; , \\
		\sigma^i_{(2)} &=& - \frac{1}{\rab} \left[ \delta_{ij} - N^i_{AB} N^j_{AB} \right] \Delta x^j (\Delta t, x^j_B, \sigma^j) \;  , \qquad  
	\eea
\end{subequations} 
where we used the property $\sigma^i \sigma^i = 1$. Using Eq.~\ref{eq:relappC} and after some algebra, we find the following relation
\bea \lb{eq:devprodvecKK}
\fl	\frac{(\bm \sigma \times \bm R_{PA} \times \bm g_P)^i}{ g_P R_{PA} - (\bm g_P \cdot \bm R_{PA})} &-& \frac{(\bm \sigma \times \bm R_{PB} \times \bm g_P)^i}{ g_P R_{PB} - (\bm g_P \cdot \bm R_{PB})}  = \\
\fl	&& \frac{ g_P (\bm \sigma \times \bm R_{PB} \times \bm g_P)^i }{ R_{PA}^2 R_{PB}^2 - (\bm R_{PA} \cdot \bm R_{PB})^2 }  \left[ R_{PA} - R_{PB} - g_P R_{AB} \right] \; .\nn
\eea
Substituting for $\Delta x^i$ from Eq.~\eqref{eq:deltax} into Eq.~\eqref{eq:sigmai} with the relation given in Eq.~\eqref{eq:devprodvecKK}, we obtain
\bea \lb{eq:sigma}
\fl	\sigma^i &=& N_{AB}^i - \frac{2G}{c^{2}} \sum_P \frac{\mathcal{M}_P}{R_{AB}} \left[ \frac{(\bm N_{AB} \times \bm R_{PB} \times \bm g_P)^i}{ g_P^2 R_{PB}^2 - ( \bm{g_P} \cdot \bm R_{PB})^2} \big(g_P R_{PA} - g_P R_{PB} - g_P^2 R_{AB} \big)  \right. \nn \\
\fl	&& \qquad\qquad \left. + (g_P^i - \nab^i \nab^j g_P^j) \ln \left( \frac{g_P R_{PB} - \bm g_P \cdot \bm R_{PB}}{g_P R_{PA} - \bm g_P \cdot \bm R_{PA}} \right)  \right] + {\cal O} (c^{-4}) \; .
\eea
It is then straightforward to check that Eq.~\eqref{eq:ki_monopmvmt} is equivalent to Eq.~\eqref{eq:reldotxhatk} when using Eq.~\eqref{eq:dotx}, Eq.~\eqref{eq:deltaxi}, Eq.~\eqref{eq:sigma} and the metric tensor~\eqref{IAUmetric} at reception event.

%%%%%%%%%%%%%%%%%%%%%%%%%%%%%%%%
\subsection{RAMOD modeling} \lb{sec:RAMOD}
%%%%%%%%%%%%%%%%%%%%%%%%%%%%%%%%
Based on a fully dynamical post-Minkowskian background~\cite{2011CQGra..28w5013C}, RAMOD has been solved explicitly in the 1PM static approximation~\cite{2013arXiv1305.4824C} needed for GSR. 
RAMOD always relies on measurable quantities with respect to a local barycentric observer along the light ray~\cite{2004ApJ...607..580D}. The unknown is the local line-of-sight, quoted $\bm{\bar \ell}$ in RAMOD and measured by the fiducial observer $\bm u$ along the null-geodesic
\be \lb{eq:def_ramod}
	\bar \ell^\alpha = - \frac{k^\alpha}{u_\beta k^\beta} - u^\alpha \; ,
\ee
where $k^\mu$ represent the tangent vectors. In this formalism, the null-geodesic equation transforms, according to the measurement protocol procedure, into a set of coupled nonlinear differential equations, called "master equations"
\begin{subequations}
\bea 
\fl	 && \frac{d\bar\ell^0}{d\zeta}- \bar\ell^i \bar\ell^j h_{0j,i} -\frac{1}{2}  h_{00,0} =0 \; ,\\
\fl	 && \frac{d\bar\ell^k}{d\zeta}-\frac{1}{2} \bar\ell^k\bar\ell^i \Big( \bar\ell^j h_{ij,0} - h_{00,i} \Big) +\bar\ell^i\bar\ell^j\left( h_{kj,i}-\frac{1}{2} h_{ij,k}\right) \nn \\
\fl	 && \qquad + \bar\ell^i \Big( h_{k0,i} + h_{ki,0} - h_{0i,k} \Big) -\frac{1}{2} h_{00,k} -\bar \ell^{k} \bar \ell^{i} h_{0i, 0} + h_{k0,0} =0 \; , \label{eq:diffeqk}
\eea
\end{subequations}
where $\zeta$ is a parameter along the null-geodesic.
Comparisons between RAMOD and other PM/PN astrometric models can be found in~\cite{2011CQGra..28w5013C}, where the author shows how RAMOD master equations recover the analytical linearized case used in~\cite{1999PhRvD..60l4002K} once converted in a coordinate form while in~\cite{2010A&A...509A..37C} the authors present a study  of the aberration in RAMOD and GREM. 
An analytical cross-check of the time of flight and tangent vectors has not been done yet with the TTF. We perform it in the static case, $i.e.$ in the case of a fully analytical solution~\cite{2013arXiv1305.4824C,2013CQGra..30d5009B}, described by the gravitational perturbation
\bea \lb{staticmetric}
\fl \qquad	h_{00}&=& \frac{2 G}{c^2} \sum_P \frac{{\cal M}_P}{r_P(\zeta)} \; , \qquad \qquad h_{0i}= 0 \; , \qquad\qquad  h_{ij}= \delta_{ij} \, \gamma \, h_{00} \; ,
\eea
where $\bm r_P(\zeta) = \bx (\zeta) - \bx_P (t_C)$ is the distance between the positions of the photon $\bx (\zeta) = \bx_B - \zeta \bm \rab$ and of the deflecting body $\bx_P (t_C)$.
%%%%%%%%%%%%%%%%%%%%%%%%%%%%%%%%%%%%%%
\subsubsection{Time of flight.}
The computation of the coordinate time of flight $ \Delta t$ can be obtained within RAMOD by considering the time component of the fiducial observer $\bm u$~\cite{2006ApJ...653.1552D}
\begin{equation} \lb{eq:u0}
 u^0 \equiv \frac{c dt}{d  \zeta}= 1 + \frac{h_{00}}{2} + O (c^{-4}) \; . 
\end{equation}
Inserting Eq.~\eqref{staticmetric} into Eq.~\eqref{eq:u0} and integrating between the emission $\zeta_A$ and the reception $\zeta_B$, we get
\begin{eqnarray}
\fl	c \Delta t &=& \int_{ \zeta_A}^{ \zeta_B} \left(1 + \frac{G}{c^2}  \sum_{P} \mathcal{M}_P \frac{1}{r_P(  \zeta)} \right) d   \zeta + {\cal O} (c^{-4})  \nn \\
\fl   			&=&  \Delta \zeta + \frac{G}{c^2}  \sum_{P} \mathcal{M}_P \ln {  \frac{\rpb + \bm \nab \cdot \bm \rpb}{\rpa + \bm \nab \cdot \bm \rpa} } + {\cal O} (c^{-4}) \; , \lb{Deltat}
\end{eqnarray} 
where $\Delta  \zeta \equiv  \zeta_B -  \zeta_A$ and we used definitions~\eqref{eq:RPdefs}-\eqref{eq:rel_R_P} with $\bbetap=0$.
We need now an explicit expression for $\Delta  \zeta$. First, we rewrite following our notation Eq.~(18) of~\cite{2013arXiv1305.4824C}
 \begin{eqnarray} \lb{barelltau}
\fl \bar{\ell}^{k}_B &=&  \frac{x^k_B - x^k_A}{\Delta \zeta} + \frac{2G}{c^{2}} \sum_{P} \mathcal{M}_P \left \{ \frac{\nab^k}{2} \left[\frac{1}{\Delta  \zeta} \ln \left[\frac{\bm \nab \cdot \bm \rpb +\rpb }{\bm \nab \cdot \bm \rpa +\rpa} \right] - \frac{1}{ \rpb } \right] \right.  \nonumber \\
\fl & &\qquad \qquad  \left. +  \frac{d^k_B }{d^2_B} \left[ - \frac{\bnab \cdot \bm \rpb }{\rpb }   +\frac{\rpb - \rpa}{\Delta \zeta }\right]  \right\}+  O \left( c^{-4} \right) \; 
\end{eqnarray}
with $\bm d_B = \brpb - \bnab (\brpb \cdot \bnab)$.
Then, using the relation $\bm d_B \cdot \bnab = 0$ and the normalisation condition $\bar \ell^\alpha \bar \ell_\alpha = g_{\alpha \beta} \bar \ell^\alpha \bar \ell^\beta = 1$ on Eq.~\eqref{barelltau} we obtain 
\bea \lb{ellkellk}
\fl \bar{\ell}^{k}_B \bar{\ell}^{k}_B &=&  1 - h_{00} \vert_B + {\cal O}(c^{-4}) = \frac{\rab^2}{\Delta  \zeta^2} + \frac{2G}{c^{2}} \sum_{P} \mathcal{M}_P \\
\fl && \qquad\qquad \times \left \{  \frac{\rab}{\Delta  \zeta^2} \ln \left[\frac{(\bnab \cdot \bm \rpb) +\rpb }{(\bnab \cdot \bm \rpa) +\rpa} \right] - \frac{\rab}{ \rpb \Delta \zeta}  \right\}+  {\cal O} \left( c^{-4} \right) \; .\nn 
\eea
Following Eq.~\eqref{Deltat}, we assume that $\Delta \zeta$ admits a PN expansion
\be \lb{eq:hattauPN}
\Delta  \zeta = \rab + \Delta  \zeta_{(2)} + {\cal O}(c^{-3}) \; ,
\ee
where $\Delta  \zeta_{(2)}$ is of order ${\cal O} (c^{-2})$.
Substituting for $\Delta  \zeta$ from Eq.~\eqref{eq:hattauPN} into Eq.~\eqref{ellkellk} and identifying the terms of the same order, we get straightforwardly
\be \lb{eq:hattau2}
\Delta  \zeta_{(2)} = \frac{G}{c^{2}} \sum_{P} \mathcal{M}_P \ln \left[\frac{\rpb + \bnab \cdot \bm \rpb }{\rpa + \bnab \cdot \bm \rpa} \right] \; .
\ee
Finally, substituting for $\Delta  \zeta$ from Eq.~\eqref{eq:hattauPN} and Eq.~\eqref{eq:hattau2} into Eq.~\eqref{Deltat} we retrieve the Shapiro term of Eq.~\eqref{ttf_ordre1} with $\bbetap = 0$.

%%%%%%%%%%%%%%%%%%%%%%%%%%%%%%%%%%%%%%
\subsubsection{Tangent vectors.}
The relation between the tangent vectors $\hat k_i$ of the TTF formalism and the the local line-of-sight $\bar \ell^i$ is obtained by expanding Eq.~\eqref{eq:def_ramod} with the metric~\eqref{staticmetric} and Eq.~\eqref{eq:u0}, so that
\be \lb{ramod_def}
	\bar \ell^i = - \frac{ k^i }{u^0 k_0} = - \hat k_i \left[ 1 - \frac{3}{2} h_{00} \right] + {\cal O}(c^{-3}) = - \hat k_i \left[ 1 - \frac{3 G}{c^2} \sum_P \frac{{\cal M}_P}{r_P(\zeta)} \right] + {\cal O}(c^{-3}) \, . 
\ee
Substituting for $\bar \ell^i$ from Eq.~\eqref{barelltau} into Eq.~\eqref{ramod_def} and using Eq.~\eqref{eq:hattauPN}-\eqref{eq:hattau2}, the reader can easily retrieve Eq.~\eqref{eq:ki_monopmvmt} with $\bbetap = 0$. 

%%%%%%%%%%%%%%%%%%%%%%%%%%%%%%%%%%%%%%%%%%%%
\section{Conclusions}\lb{sect:concl}
%%%%%%%%%%%%%%%%%%%%%%%%%%%%%%%%%%%%%%%%%%%%
This paper provides the integral form for the time of flight of a photon between two points at finite distance and its tangent vectors. It is remarkable that Eq.\eqref{Tr1PN} and Eqs.~\eqref{eq:kCF}-\eqref{eq:dDr1PM} give these quantities in closed form as function of just few parameters and for any metric tensor describing a weak gravitational field up to the first post-Newtonian approximation.  
We show an application in the case of the time-dependent metric tensor~\eqref{IAUmetric}, well suited for representing light propagation within the Solar System for ongoing space experiments.
Eqs.~\eqref{ttf_ordre1}-\eqref{ttf_ordre1_e} and Eq.~\eqref{eq:ki_monopmvmt} extend the results previously obtained in~\cite{2008PhRvD..77d4029L} and are used here to find a procedure to relate and compare three independent approaches to relativistic light propagation.
Sections~\ref{sec:KK92} and \ref{sec:RAMOD} show that the results of TTF and KK92 are equivalent at the first post-Newtonian approximation in a time-dependent gravitational field while the results of TTF and RAMOD are equivalent at least at the approximation required for the Gaia mission. 
Such a cross-checking procedure enters the same thread of model comparison started in~\cite{2010A&A...509A..37C}, and it is essential to fully understand the observational data coming from Gaia in a common experimental context. 

\ack
The authors are grateful to P. Teyssandier for fruitful discussions and suggestions.

\noindent S. Bertone is Ph.D. student under the UIF/UFI (French-Italian University) program and thanks UIF/UFI for the financial support of this work.
\noindent S. Bertone, C. Le Poncin-Lafitte and M.-C. Angonin are grateful to the financial support of CNRS/GRAM and CNES.
\noindent The work of M. Crosta and A. Vecchiato has been partially funded by ASI under contract to INAF I/058/10/0 (Gaia Mission - The Italian Participation to DPAC).

%%%%%%%%%%%%%%%%%%%%%%%%%%%%%
\appendix
%%%%%%%%%%%%%%%%%%%%%%%%%%%%%
\section{Numerical impact of the approximation on the perturbing body trajectory} \lb{sec:approxtrajP}

Let us expand the expression of the perturbing body $P$  orbit around $t_{C}$ as follows
\be
\bx_P(t)=\bx_P(t_{C})+c(t-t_{C}) \frac{\vp}{c}+\frac{c^2(t-t_{C})^2}{2} \frac{\app}{c^2} + ... \, , 
\ee
where we can analyse the amplitude of $\vert \vp/c\vert$ and $\vert\app/c^2\vert$ for quasi-circular orbits. Indeed we have 
\bea
&&\abs{\frac{\vp}{c}} \lesssim \sqrt{\epsilon_P} \sqrt{\frac{GM_S}{c^2~R_{PS}}},~~~\abs{ \frac{\app}{c^2}} \lesssim \epsilon_P \frac{GM_S}{c^2~R^2_{PS}},
\eea
where $M_S$ is the mass of the Sun, $R_{PS}$ the distance between the pertubing body and the Sun and $\epsilon_P =(1+e_P)/(1-e_P)$, $e_P$ being the eccentricity of body $P$. Let us consider the circular case $\epsilon_P=1$ with $\frac{GM_S}{c^2} \sim 1.5$km. If we choose $R_{PS}=10^8 km$ we get
\bea \lb{eq:ordresvitacc}
&&\abs{\frac{\vp}{c}} \sim 10^{-4},~~~\abs{ \frac{\app}{c^2}} \sim 1.5~ 10^{-16} km^{-1}.
\eea

%%%%%%%%%%%%%%%%%%%%%%%%%%%%%%%%%%
\section{Choosing $t_{C}$ and its numerical impact} \lb{sec:tDprob}

Since in astrometry, we only have direct access to the reception time $t_B$, the simplest choice would seem to set $t_{C} \equiv t_B$. Unfortunately, this choice would lead to errors in the data analysis. To illustrate this, let us assume that the reception is done on a remote spacecraft at $10^9$~km from the perturbing body.

If $t_{C}$ is defined such that $t_{C} \equiv t_B$, one has $c(t_m-t_{C}) \sim 10^9$ km, where $t_m$ is the coordinate time of the closest distance of the photon to the perturbing body. Therefore, using the orders of magnitude of Eq.~\eqref{eq:ordresvitacc}, we deduce that
\bea
&&c(t_m-t_C)\abs{\frac{\vp}{c}} \sim 10^{5} km,~~\frac{c^2(t_m-t_C)^2}{2} \abs{ \frac{\app}{c^2}} \sim 70 km ~ \nonumber \;,
\eea
meaning that by neglecting the acceleration term in the development one would have an error of $70$~km on the impact parameter ($b$) of the trajectory of the light beam in the worst case. For the deviation angle $\alpha$, one has $\alpha \propto 1/b$. Therefore, the relative error on the deviation angle will be of the order of $70~km/b$ (since $1/(b \pm 70) \sim  (1 \mp70/b)~1/b$). If $b \sim 3 \times 10^3$km (if the photon grazes Mercury for instance), then the error on the deviation angle will be around $2\%$ -- which is unnecessarily big. 

Indeed, let us compute the angular error introduced on the observation of a light signal grazing Jupiter. We can define the angular error as 
\be
\Delta \alpha = - \frac{4 G M_J}{c^2 b^2} \Delta b \; ,
\ee
where $M_J$ is Jupiter mass, Jupiter Schwarzschild radius is approximately $2.8 \; m$, $b$ is Jupiter equatorial radius for a grazing photon and $\Delta b ~=~ 70 km$ is the error on the impact parameter. Then $\Delta \alpha \sim 16 \; \mu as$, well above the desired precision for the model. The same computation for Saturn and Mars gives respectively $15 \; \mu as$ and $0.2 \; \mu as$.
%(Let us also note that if the remote saltellite is at $10^{10}$km (imagining that the mission would follow the path of Pioneer for instance), then one would have an error of $20\%$).

On the contrary, if $t_{C} \approx t_m$, then  $c(t_m-t_{C}) \sim 0$ and the error introduced by the approximation on the trajectory is small. This means that $t_{C}$ is chosen as the maximum approach time of the photon to the perturbing body, such that $|\bx_\gamma(t_{C})-\bx_P(t_{C})|\sim b$, where $\bx_\gamma(t)$ is the trajectory of the photon.

In that case, the additional $\beta c^{-2}$ terms in the time transfer or deviation angle coming from the development of the trajectories of the bodies in the application of the TTF will be at the same numerical level that the terms coming from the $c^{-3}$ part of the metric (also due to the motion of the perturbing bodies: $c^{-3} \equiv \beta c^{-2}$). 

The last statement works for the most general case and therefore one should define  $t_{C}$ as being such that $|\bx_\gamma(t_{C})-\bx_P(t_{C})|\sim b$, similarly to what stated in \cite{2003A&A...410.1063K,2003AJ....125.1580K}.

%%%%%%%%%%%%%%%%%%%%%%%%%%%%%%%%%%%%%%%%%%
\section{Emission/Reception TTF equivalence} \lb{appendix1}
In section 5, we derived expressions in Eq.~(\ref{ttf_ordre1}) and Eq.~(\ref{ttf_ordre1_e}) for the delay functions $\Delta_{r}^{(1)}$ and $\Delta_{e}^{(1)}$ in the gravitational field of point-like, slowly moving and non-rotating bodies. We prove here the formal equivalence $\Delta_{r}^{(1)} = \Delta_{e}^{(1)}$ as stated in Eq.~(\ref{ttf}). By introducing Eq.~\eqref{eq:rel_R_P} and the relation
\bea \lb{eq:relappC}
	R_{AB}^2 [g_P^2 R_X^2 - (\bm g_P \cdot \bm R_X)^2] &=& R_{AB}^2 g_P^2 R_X^2 - (R_{AB} \bm g_P \cdot \bm R_X)^2 \nn\\
				&=& R_{PA}^2 R_{PB}^2 - (\bm R_{PA} \cdot \bm R_{PB})^2 \; , 
\eea
with $X$ taking the values "$PB$" or "$PA$", it is straightforward to show that 
\bea  \lb{eq:equivAppC}
\fl	\Delta_{r}^{(1)} &=& \sum_P g_P r_P^G \ln \left[ \frac{\rab g_P \rpa - \rab \brpa \cdot \bg}{\rab g_P \rpb - \rab \brpb \cdot \bg} \right]  \nn \\
\fl		&=& \sum_P g_P r_P^G \ln \left[ \frac{\rab^2 g_P^2 \rpa^2 - \rab^2 (\brpa \cdot \bg)^2}{\rab^2 g_P^2 \rpb^2 - \rab^2 (\brpb \cdot \bg)^2} \times \frac{\rab g_P \rpb + \rab \brpb \cdot \bg}{\rab g_P \rpa + \rab \brpa \cdot \bg} \right]  \nn \\
\fl		&=& \sum_P g_P r_P^G \ln \left[  \frac{\rab g_P \rpb + \rab \brpb \cdot \bg}{\rab g_P \rpa + \rab \brpa \cdot \bg}  \right] = \Delta_{e}^{(1)} \; . 
\eea

%%%%%%%%%%%%%%%%%%%%%%%%%%%%%%%%%%

%%%%%%%%%%%%%%%%%%%%%%%%%%%%%%%%%%

%%%%%%%%%%%%%%%%%%%%%%%%%%%%%%%%%%
\bibliographystyle{iopart-num}
\bibliography{BiblioPerso} % your references in file: Yourfile.bib

\end{document}